\documentclass[epsf,usegraphicx,usenatbib]{mn2e}

\usepackage{xspace}
\usepackage{aas_macros}
\usepackage{subfigure}
\usepackage{amsmath}
\usepackage{amsfonts}
\usepackage{times}
\usepackage{url}
\title[\emph{Kepler} observations of a background dwarf]{Serendipitous \emph{Kepler} observations of a background dwarf nova of SU UMa type}

\author[Barclay et al.]
{
\parbox{\textwidth}{
Thomas Barclay$^{1,2}$\thanks{E-mail: thomas.barclay@nasa.gov}, 
Martin Still$^{1,2}$, 
Jon M. Jenkins$^{1,3}$,
Steve B. Howell$^{1}$,
Rachael M. Roettenbacher$^{4}$
}
\vspace*{4pt}\\
$^{1}$ NASA Ames Research Center, M/S 244-40, Moffett Field, CA 94035, USA\\ 
$^{2}$ Bay Area Environmental Research Institute, Inc., 560 Third St. West, Sonoma, CA 95476, USA\\
$^{3}$ SETI Institute/NASA Ames Research Center, Moffett Field, CA 94035, USA\\
$^{4}$Deptartment of Astronomy, University of Michigan, 830 Dennison Building, 500 Church Street, Ann Arbor, MI 48109, USA
}
\newcommand{\K}{{\sl Kepler} }

\begin{document}

\date{\today} 
\maketitle

\maketitle

\begin{abstract}

We have discovered a dwarf nova (DN) of type SU UMa in \K data which is 7.0 arcsec from the G-type exoplanet survey target KIC 4378554. The DN appears as a background source in the pixel aperture of the foreground G star. We extracted only the pixels where the DN is present and observed the source to undergo five outbursts -- one a superoutburst -- over a timespan of 22 months. The superoutburst was triggered by a normal outburst, a feature that has been seen in all DNe superoutburst observed by \emph{Kepler}. Superhumps during the super outburst had a period of $1.842\pm0.004$ h and we see a transition from disc-dominated superhump signal to a mix of disc and accretion stream impact. Predictions of the number of DNe present in \K data based on previously published space densities vary from 0.3 to 258. An investigation of the background pixels targets would lead to firmer constraints on the space density of DN.

\end{abstract}


\begin{keywords}
\end{keywords}

\section{Introduction}

Cataclysmic variables (CVs) are binary star systems consisting of a white dwarf primary and a less massive main sequence or post-main sequence secondary.  The two stars orbit a common centre of mass with typical periods from between 50 min and 5.7 d \citep{downes01,ritter03,thorstensen06,sekiguchi92}. The small physical size of the binary system causes the secondary to be distorted by the gravitational field of the white dwarf and leads to the filling of its Roche lobe. Mass is transferred through the L1 inner Lagrangian point and forms a stream of material falling into the potential well of the white dwarf. In the case where the white dwarf is non-magnetic, the infalling material forms an \emph{accretion disc} \citep{shakura73}. Viscous shearing layers amplify a weak magnetic field within the disc, and lead to turbulent angular momentum transport radially outwards \citep{balbus98}. Mass can then move inwards and be accreted onto the surface of the white dwarf \citep{hellier01}. At optical wavelengths, the disc is usually much more luminous than the white dwarf, its companion, and the impact of the accretion stream onto the disc.

The DN subclass of CVs undergo quasi-periodic outbursts \citep[see][for a review]{warner95}. During an outburst the optical brightness of the system can increase by a factor of 10 or more \citep[e.g.][]{ramsay09}, which interrupts the faint \emph{quiescent} state in which the DNe typically spend most of their time.

If the mass transfer rate from the secondary to the accretion disc is greater than the rate of disc accretion onto the white dwarf, $\dot{M}(2) > \dot{M}(d)$, then the disc radius will increase over time. This can only happen if the dimensionless disc viscosity parameter $\alpha$ is low \citep[where low is usually taken in models to be  $\la 0.03$, e.g.][]{smak84,cannizzo93}.  The disk instability model predicts that eventually the surface density in the disc will exceeds a critical value $\Sigma_{\textrm{max}}$, which initiates a phase transition in the disc material \citep{hoshi79,lasota01}. At this point the $\alpha$ parameter increases \citep[by a factor of $\sim 10$][]{smak84,cannizzo93} and $\dot{M}(2) < \dot{M}(d)$ for a time until the surface density at the outer edge of the disc decreases below some critical surface density $\Sigma_{\textrm{min}}$, and $\alpha$ returns to the original lower value. This model is known as the limit cycle accretion disc instability \citep{lasota01}.




A subset of DNe known as SU UMa stars occasionally undergo a superoutburst (SO) in addition to undergoing normal outbursts (NO). SOs are typically a factor of two brighter and 5--10 times longer than a NO \citep{warner95}. Superhumps,  photometric variability on a period a few per cent longer than the orbital period of the system, are observed during the SO. During a NO the disc expands in radius \citep{osaki89}. After a number of these outbursts, the disc can become large enough to grow beyond the inner Lindblad resonance, close to the 3:1 co-rotation resonance with the binary orbital period\citep{whitehurst88}. Once there is enough mass at this resonance radius, the superhump oscillation can be driven by tides.

We identified a background source in \K spacecraft archival data whose light curve resembled that of a DN. In \S\ref{sec:obs} we describe the observations made by the \K spacecraft of this source, our re-extraction of the light curve from pixel level data using a more appropriate pixel aperture, and our removal of systematic photometric effects from the flux time series. In \S\ref{sec:outbursts} we discuss observed outbursts, and in \S\ref{sec:disc} we discuss the implications of these observations on the nature of outbursts from DN.

\section{Observations}
\label{sec:obs}
\K has been almost continuously observing over $1.6\times 10^5$ stars at a cadence of 30 min \citep{borucki10,koch10} at milli- to micro-magnitude precision \citep{gilliland11} since the spacecraft's launch on 2009 March 06. While the primary aim of the \K mission is the discovery of exoplanets, the long baseline of continual, high precision observations has also provided an unrivalled data-set to the stellar astrophysics community.

The available bandwidth does not allow all starlight falling onto the \K detectors to be downloaded at a cadence of 30 min. Instead the pixels to be collected must be preselected and account for only 5.75 per cent of the total detector area available \citep{koch10}. 

The continuity of observations are broken every month by the satellite pointing towards Earth for data to be downloaded. Once every three months the satellite rolls $90^{\circ}$ to keep the sunshade pointed in the appropriate direction but has the effect of changing the CCD that a star falls upon every quarter. Because of the change in physical position of the instrument and because a new data-set is downloaded, the data from \K is divided into quarters starting with Quarter 0 (Q0), containing commissioning observations, and full science operations beginning in Q1.

Each observed star is extracted by selecting a pixel mask which can change in size and shape every quarter because the position of the star in pixel space is different. For this reason the contamination from background stars is different every quarter -- this contamination can be significant because part of the \K  field covers the Galactic plane (where the stellar density is high) and the pixels are relatively large ($3.98''\times 3.98''$).

The observations described in this paper are of one particular target aperture -- that around KIC 4378554. This star was chosen to be a target in the \K exoplanetary program. \K observed KIC 4378554 during Q1--8 excluding Q6 (see Table~\ref{tab:observations} for a log of the observations). No planet has been found transiting this system as yet. The \K Input Catalogue lists KIC 4378554 as a $K_{p} = 14.9$ star\footnote{$K_{P}$ refers to approximate Kepler instrumental magnitudes. For more details of this passband see \citet{brown11}.} with colour $(g'-r') = 0.57$, an effective temperature of 5617 K, and a surface gravity of $\log g = 4.615$ in cgs units; these values are consistent with the source being a main sequence G-type star \citep{brown11}. The star is only visible during three of the four seasons of observations because it lies on Module 3 for one quarter. Module 3 failed during Q4.


\begin{table*}
\begin{center}
\caption{Log of \emph{Kepler} observations of KIC 4378554 up to BJD 2455635. The start and end Barycentric Julian Date (BJD) refer to the mid-time of the first and last cadences on which there are no flags questioning the quality of the data. We find the data quality flags in the light curve and target pixel files under the column SAP\_QUALITY -- non-zero indicates a potentially compromised cadence \citep[the specific meaning of each flag is given in the Kepler Archive Manual][]{archivemanual}. Likewise, the number of cadences column in this table excludes cadences which contain potentially compromised data.\label{tab:observations}}
\begin{tabular}{crrrrrr}
\hline\hline
Quarter & Start BJD & End BJD & \# cadences & Module & Output & Channel\\
& $-2454900$& $-2454900$\\
\hline
1 &64.5115 &97.9832 & 1611 & 15 & 3 & 51\\
2 &102.7648 &191.4672 & 4048 & 3 & 3 & 7\\
3 &193.2449 &282.4947 & 4099 & 11 & 3 & 35\\
4 &285.3758 &375.2017 & 4092 & 23 & 3 & 79\\
5 &376.5095 &471.1621 & 4460 & 15 & 3& 51\\
7 &563.1953 &652.5473 & 4190 & 11 & 3& 35\\
8 &668.3827 &735.3435 & 3098 & 23 &3 & 79\\
\hline
\end{tabular}
\end{center}
\end{table*}

\subsection{Optimal aperture observations}

\begin{figure*}
\includegraphics[width=\textwidth]{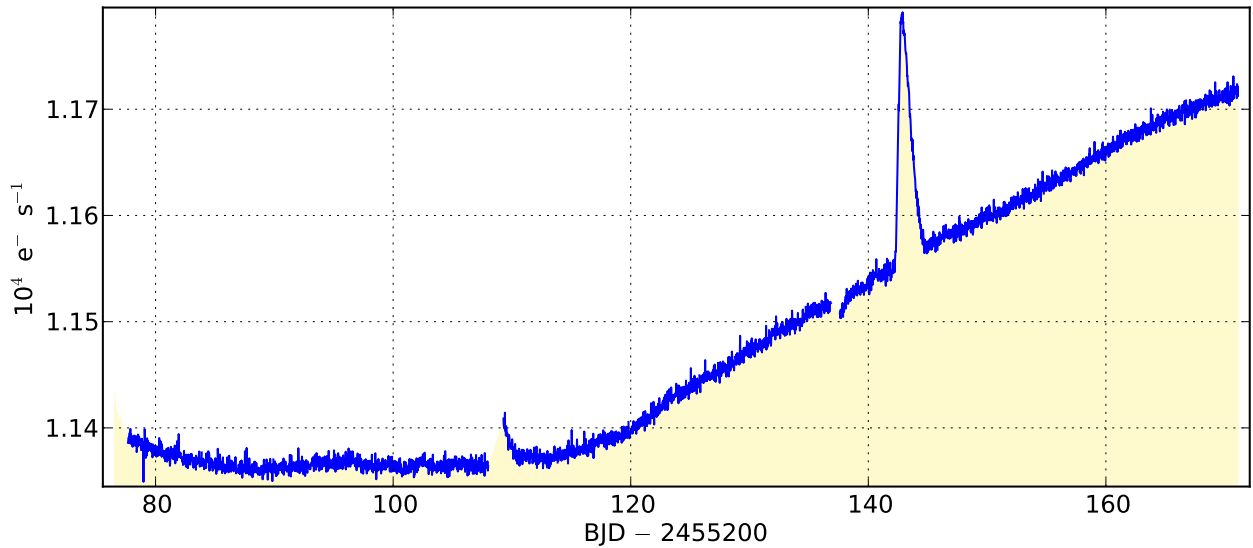}
\caption{The Q5 Simple Aperture Photometry (SAP) light curve made using the optimal aperture calculated for KIC 4378554. The light curve contains systematics, such as differential velocity aberration, which causes the change in the general flux level over the last 50 days of the quarter, and a data-gap from Earth-point followed by a thermal settling at BJD 2 455 307 and then again at BJD 2 455 337. In addition, there is a sudden, temporary 2 per cent increase in the flux level, reaching a peak at BJD 2 455 342.8. \label{fig:saplc}}
\end{figure*}

Two light curve products are provided by the \K project: light curves extracted using Simple Aperture Photometry (SAP) and light curves which have undergone the additional step of Pre-Search Data Conditioning \citep[PDC,][]{twicken10}. The latter of these is produced for use in exoplanet searches but is known to cause problems if used for stellar astrophysics work. These problems include the attenuation of astrophysical signal and the injection of spurious signals, because of this we use the SAP data. A light curve showing Q5 data can be seen in Fig~\ref{fig:saplc}. These data have not had systematic artefacts removed, and there are several noticeable features due to systematics such as the increase in flux level over the final 50 d of the quarter and the thermal settling after Earth-points at BJD 2 455 307 and 2 455 337. However, there is one obvious feature that appears to be astrophysical in nature -- a 2 per cent increase in brightness with a peak at BJD 2 455 342.8 lasting 2.5 d. In addition to the brightening shown here, we identified four similar events -- two in Q3, one in Q4 and one in Q7. Events like these are not typically seen in G-type stars and therefore warranted further investigation. 

\subsection{Custom light curve extraction from target pixel files}
\begin{figure*}
\includegraphics[width=\textwidth]{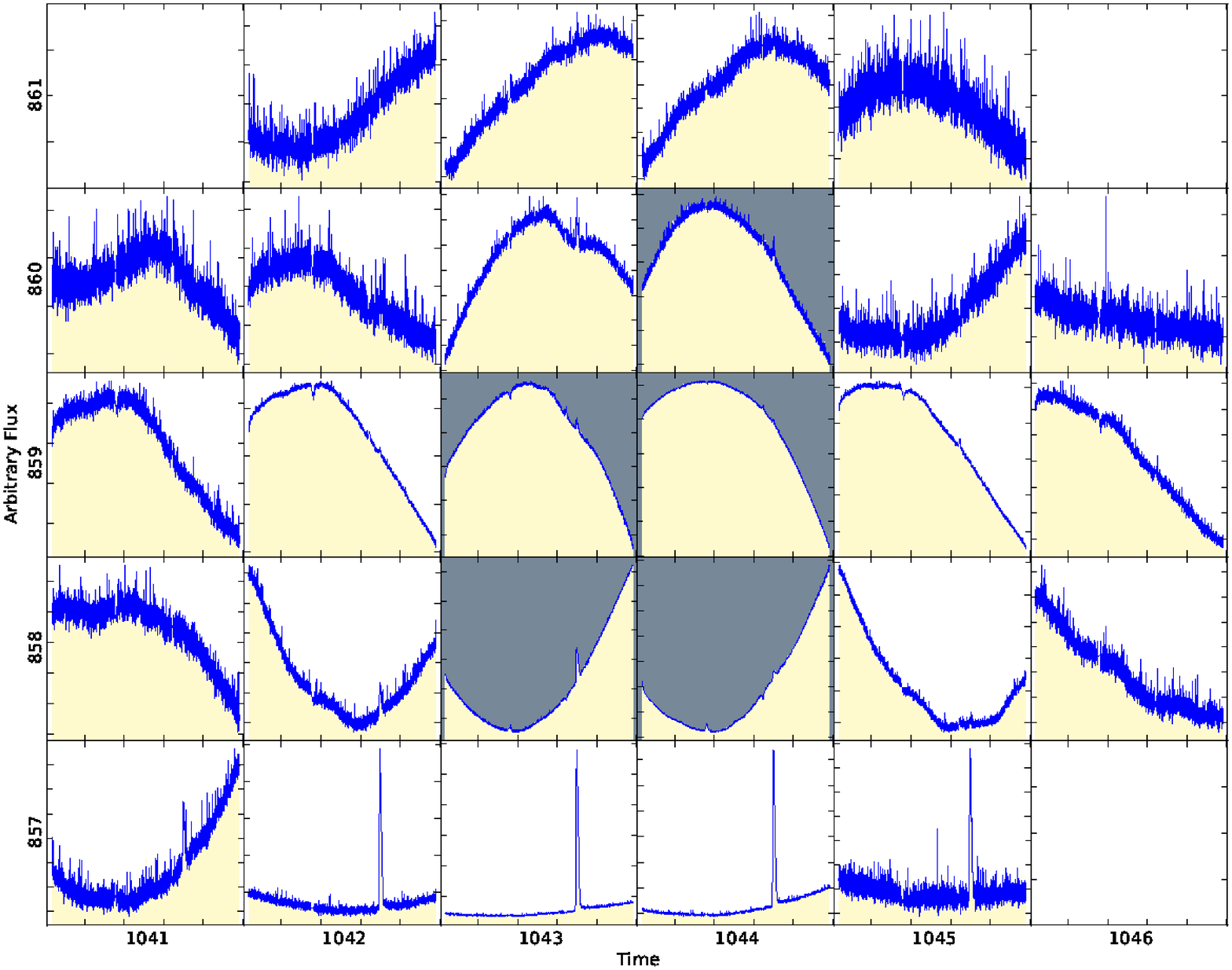}
\caption{Plotted is the flux time series within each pixel collected for KIC 4378554 \textbf{during Quarter 5, where a Quarter is the number of days between each spacecraft roll. The quarter length is typically 91 days but can vary because of, for example, spacecraft anomalies.} The numbering around the plot provides the pixel column and row on the CCD channel. The blank boxes contain pixels that were not within the pixel mask. The five boxes with the grey background are those within the optimal aperture of KIC 4378554 and were used to make the plot shown in Fig~\ref{fig:saplc} and the archived light curve at the Multimission Archive at STScI (MAST). The brightening event is not coming from the source within the optimal aperture but from another source on the bottom edge of the pixel mask. The flux range is scaled separately for each pixel, the amount of scatter is in each pixel is indicative of the signal-to-noise of the data in that pixel. Differential velocity aberration causes KIC 4378554 to move towards the lower left corner over the length of the quarter. The thermal settlings cause KIC 4378554 to move from left to right.\label{fig:pixseries}}
\end{figure*}

In order to determine whether the brightness increases are due to KIC 4378554 or a background star in the optimal aperture of KIC 4378554, we examined the archived target pixel files\footnote{Kepler target pixel and light curve files are distributed at \url{http://archive.stsci.edu/kepler}.}. These files contain a series of images -- one for each time-stamp --  showing the individual pixels downloaded from the \K spacecraft. These images were calibrated by the \K pipeline \citep[as described by][]{quintana10}. Shown in Fig.~\ref{fig:pixseries} is the Q5 time-series for each calibrated pixel in the aperture of KIC 4378554. The pixels shown in grey were used to create the light curve in Fig~\ref{fig:saplc}. From  Fig.~\ref{fig:pixseries} we can see that the brightness increase at BJD 2 455 342.8 is not caused by KIC 4378554 but from a source which appears to have its centre on or close to pixel (1043,857) approximately 1.5 pixels south-west of the target star. For convenience we will refer to this source as NIK 1 (standing for Not In KIC). 

We extracted a new light curve by summing only those pixels which NIK 1 contributes flux using the \textsc{kepmask} and \textsc{kepextract} tools from the \textsc{PyKE} \K community software\footnote{\textsc{PyKE} is available from \url{http://keplergo.arc.nasa.gov/ContributedSoftware.shtml}}. In quarters where we see a sudden brightening event, this is relatively straight forward. This is the case in Q3, 4, 5 and 7. For Q1 we use the Q5 aperture  and for Q8 we used the Q4 aperture because NIK 1 will be at the same position on the CCD four quarters apart. For Q2 we estimate the position of NIK 1 from the full-frame image taken at the end of Q3, when a brightening event occurred and NIK 1 is visible\footnote{\K full-frame images are collected once a month and are available from \url{http://archive.stsci.edu/kepler/ffi/}.}. We overlayed the full-frame image containing NIK 1 on the pixel level data and selected the appropriate pixels. The number of pixels we use for all quarters is listed in Table~\ref{tab:pixels}.

\begin{table*}
\begin{center}
\caption{Pixels used to extract the light curve of NIK 1 in the optimal and custom apertures, the time spans masked out of the cotrending basis vector fit and the number of basis vectors used in these fits. \label{tab:pixels}}
\begin{tabular}{ccccc}
\hline\hline
Quarter & Pixels in optimal ap.&Pixels in custom ap.& Time range masked in CBV fit& \# of CBVs in fit\\
&&&BJD - 2454900\\
\hline
1&6&2&&3\\
2&5&1&&3\\
3&6&5&272.71465--282.50147&5\\
4&4&2&285.35980--293.98626&5\\
5&5&2&441.89137--448.34535&6\\
7&6&5&621.90676--629.78812&6\\
8&4&2&&3\\
\hline
\end{tabular}
\end{center}
\end{table*}

\subsection{Removing systematic trends using cotrending basis vectors}
The light curves we extracted from target pixel files still contain systematic features. This is because the extracted aperture does not contain all the flux and also contains contaminating flux from the target star,  KIC 4378554. The amount of flux in the aperture, both from NIK 1 and from other contaminating sources, changes continually over time due to differential velocity aberration and suddenly due to changes in the focus after Earth-points \citep{christiansen11}. 

The instrumental systematics which affect a single light curve are unique. However, we see strong correlations between stars which are on the same CCD channel. We can take those stars which are most highly correlated, which we call the reference ensemble, and identify the correlated features. These correlations can be represented as linear combinations of orthogonal vectors, which can be fit and subtracted from each individual light curve to remove instrumental features. The \K team have created a product containing these vectors, known as cotrending basis vectors (CBVs), and have made them available to the public\footnote{CBVs are available from \url{http://archive.stsci.edu/kepler/cbv.html}.} \citep[a description of how they were created is given in][]{DRN11,DRN12}. 

The available files consist of the 16 vectors, of length the same as the light curve, which provide the greatest contribution to the variance of the reference ensemble. There is one set of 16 vectors per quarter for every CCD readout. It is crucial that an appropriate number of vectors are chosen to fit to to the light curve data; using too few will leave instrumental signal in the light curve, using too many will cause over-fitting and remove astrophysical signal.

We used the \textsc{kepcotrend} tool from \textsc{PyKE} to perform a linear least squares fit of the CBVs to the light curve we extracted using \textsc{kepextract}, and masked the regions where the brightening events occur out of the fit because these regions are dominated by astrophysical signal not systematics. We selected the number of CBVs to fit by experimenting with the fit: we assumed that there was no intrinsic long period variability outside of the brightening and attempted to remove any long term trends. Starting with two CBVs, we iteratively increased the number of CBVs until the regions outside of the brightening events was approximately constant. The number of CBVs used and the regions masked from the fit are listed in Table~\ref{tab:pixels}.

The CBVs only correct for the amplitude of systematic effects and not for constant offsets caused by differences between the contamination from nearby sources and changes in the fraction of flux of NIK 1 in the pixel aperture in different quarters. After removing the instrumental systematics there are still differences between the median flux levels between quarters. To mitigate the effects of this we normalise the light curve. A representation of the true flux level was obtained by first normalising independently for each quarter:
\begin{equation}
F_{\mathrm{norm},i}= \frac{F_{i} - \mathrm{med}\left(F\right)}{\mathrm{MAD}\left(F\right)}
\end{equation}
where $F_{\mathrm{norm}}$ is the vector of normalised fluxes, $F$ is the vector of non-normalised fluxes, $F_{i}$ is the $i^{\mathrm{th}}$ element of vector $F$, $\mathrm{med}\left(F\right)$ is the median of $F$, and $\mathrm{MAD}\left(F\right)$ is the median absolute deviation from the median (MAD) of F. MAD is defined as 
\begin{equation}
\mathrm{MAD}\left(F\right) = \mathrm{med} \left(|F_{i} - \mathrm{med}\left(F\right)|\right).
\end{equation}

\begin{figure*}
\includegraphics[width=\textwidth]{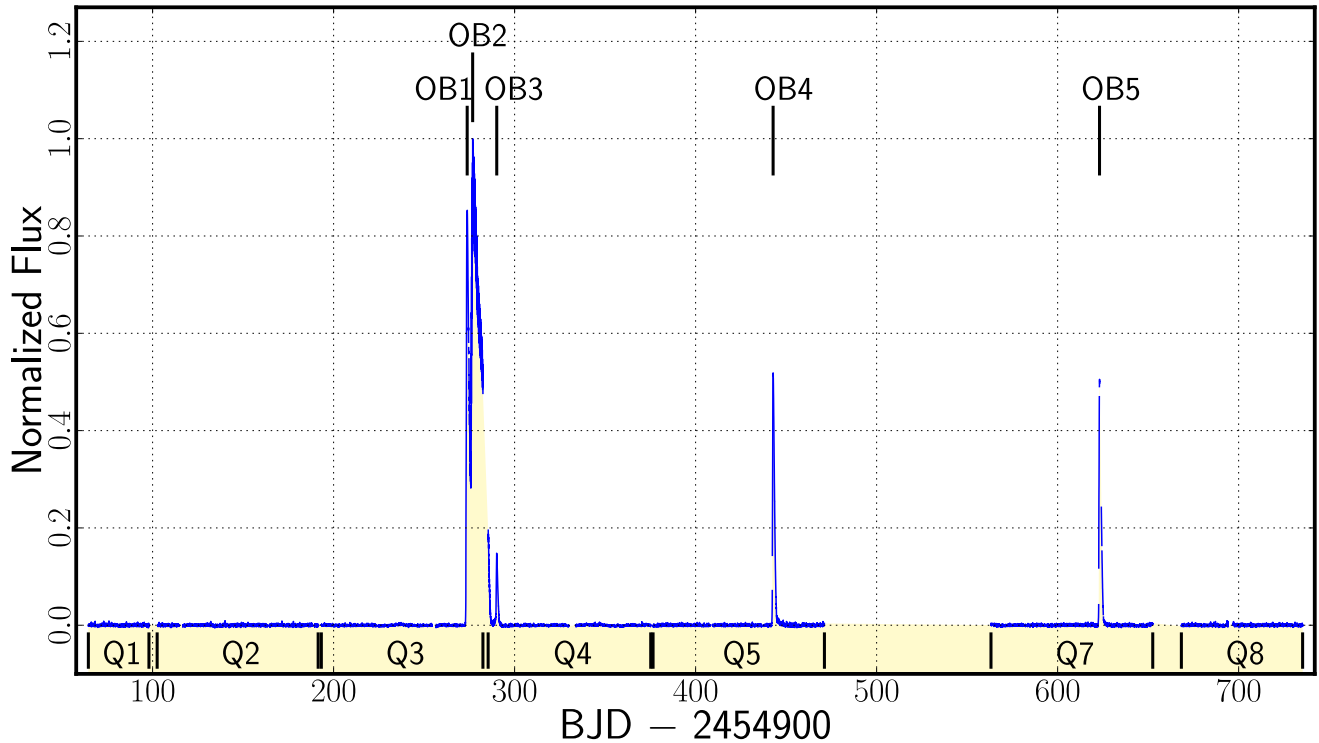}
\caption{The fully processed light curve of NIK 1 showing seven quarters of data. The flux time series has been extracted from the target pixel files, corrected by fitting and subtracting the CBV and normalised with weights. The normalisation was done by first dividing each quarter of data by its median and then dividing this value by the median absolute deviation of the corrected time series. We then divided the whole data-set by the median flux of that data. The five outbursts we observed are marked with OB1--5 and beneath the time series is shown the quarter over which the data were taken. \label{fig:fulllc}}
\end{figure*}

We show the full custom extracted and cotrended time series containing seven quarters of data in Fig.~\ref{fig:fulllc}. The light curve is that of a SU UMa type DN \citep[cf. the adapted light curve of VW Hyi from Bateson (1977) shown in][]{warner95}. We observe five outbursts (labelled OB1--5 in Fig.~\ref{fig:fulllc}): four NOs and one SO. One of the NOs (OB1) immediately precedes the SO (OB2) while another NO (OB3) is of much lower intensity than the other NOs, and occurs a mere 4 days after the end of OB2. Our method of normalisation is relatively crude; \textbf{if, for example, there are significantly different levels of contamination or the noise level was dramatically different between the quarters, we would be incorrectly estimating the relative outburst amplitudes}. We gain confidence that the normalisation is appropriate for the following reasons; the outbursts in Q5 an Q7 are of comparable intensities, and there is a relatively smooth transition from the tail of the SO seen in Q3 and into Q4. It is possible that we are severely underestimating the amplitude in Q4 and OB3 is actually of comparable intensity to OB1, OB4 and OB5, but this would require a sharp break in the decline of the SO in the data gap between Q3 and Q4.


\subsection{Archival Observations}

We were unable to locate NIK 1 in quiescence in \K full-frame images which have a limiting magnitude of $K_{P}\sim 20$, and so we looked to archival data to find NIK 1 in a quiescent state. Palomar Sky Survey I and II (POSS-I and II)  observations appear in the USNO-B1.0 catalogue \citep{monet03}. USNO-B1.0 lists a source offset (in the direction of KIC 4378554) from NIK 1 by 2.7 arcsec. The POSS-I blue magnitude of this source is given as 19.15 and  the source is catalogued as USNO-B1 1294-0347652. They again find this source in the POSS-II blue image and derive a blue magnitude of 15.13. 

\begin{figure*}
\includegraphics[width=\textwidth]{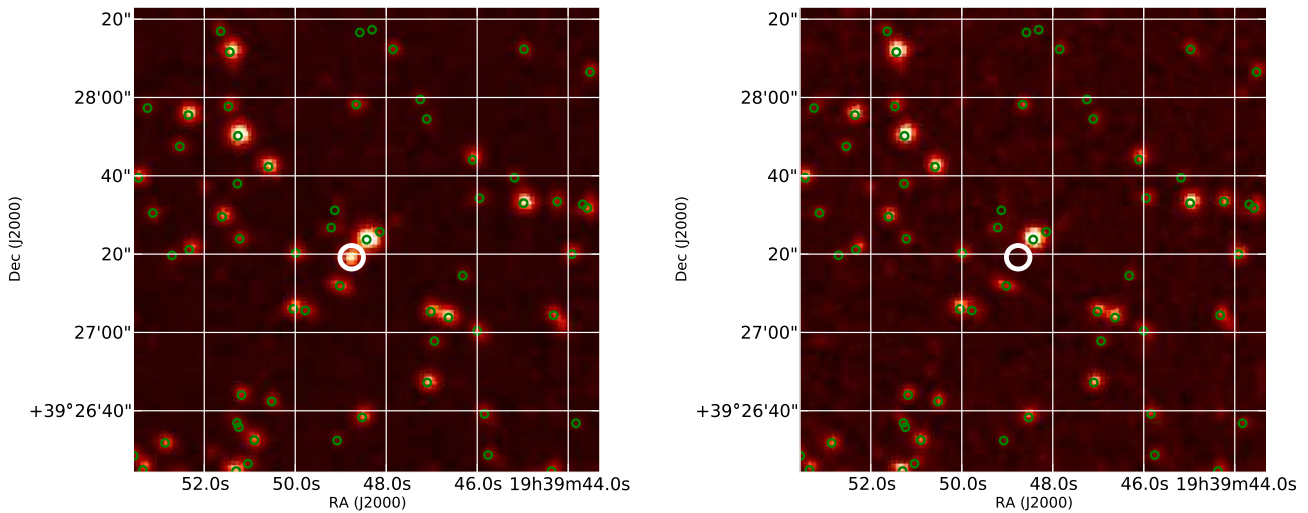}
\caption{The DSS2 blue image taken on JD 2447351.8 is shown in the left panel and the DSS2 red image taken on JD 2448508.7 in the right panel. Circled in white on the right panel is a source in the same position as we identified NIK 1 in the Kepler images. A white circle in drawn in the same position in the left panel but no source is detected. The green markers are the positions of sources which are in the Kepler Input Catalogue, no source exists in the catalogue at the position where we have detected the outbursting star. \label{fig:dss}}
\end{figure*}

POSS-I and II images are available of this region as part of the Digitized Sky Survey (DSS). We examined the POSS-II images of this region and found a source we estimate to be at $B\simeq17.3$ based on the brightness of nearby stars in the blue image, much brighter than the detection limit of the full-frame images where we do not detect NIK 1. We deduce that this source is NIK 1 in a state of outburst. Fig.~\ref{fig:dss} shows a comparison between the POSS-II red and blue images which were taken at different epochs. No source is detected at the position of NIK 1 in the red image. 

Despite the USNO-B1.0 listing a source at the same position in POSS-I blue data we were unable to locate any source. However, the fact that the source is listed with two quite different brightness measurements leaves us in little doubt that USNO-B1 1294-0347652 and NIK 1 are one and the same.



\section{Outbursts}
\label{sec:outbursts} 
The \K data contain observations of five outbursts: four NOs and one SO. The dates the outbursts reach maximum and their rise and fall durations are given in Table~\ref{tab:outbursts}. 

\begin{table*}
\begin{center}
\caption{Outburst occurrences. \label{tab:outbursts}}
\begin{tabular}{cccccc}
\hline\hline
Outburst & Quarter & Type of outburst & BJD of $F_{max}$ & $T_{rise}$ & $T_{fall}$ \\
\hline
1& 3 & Normal&2455173.85&1.08&1.96 (a)\\
2& 3 & Super &2455176.86&1.05 (b)&10.42\\
3& 4 & Normal&2455190.10&1.25&1.86\\
4& 5 & Normal&2455342.84&0.86&2.57\\
5& 7 & Normal&2455523.17&0.81&2.39\\
\hline
\multicolumn{6}{l}{(a) The fall time is truncated by the onset of the SO}\\
\multicolumn{6}{l}{(b) The rise time is truncated by the tail  of the first NO}\\
\end{tabular}
\end{center}
\end{table*}

\subsection{Normal outbursts}
\begin{figure*}
\includegraphics[width=\textwidth]{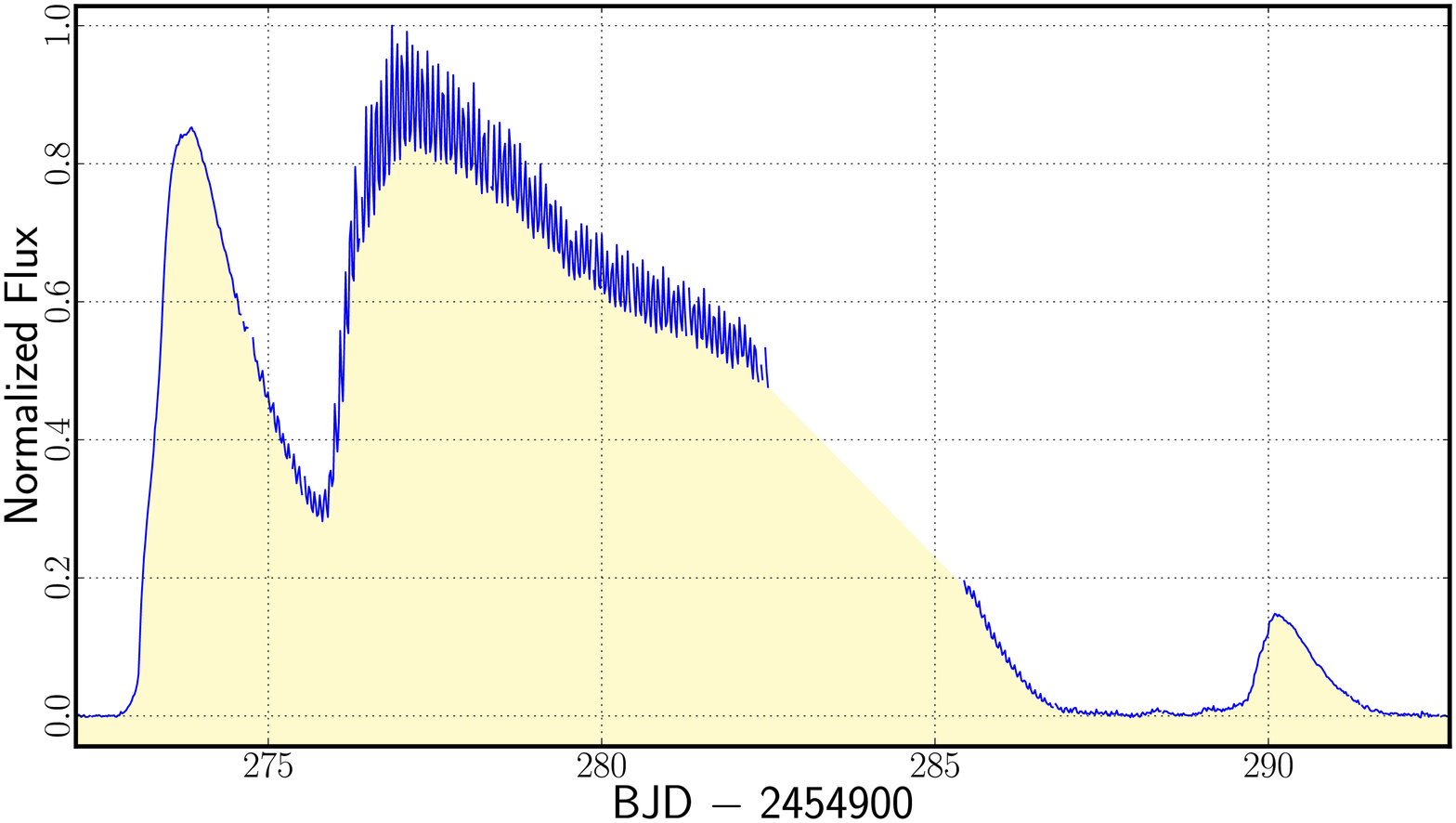}
\caption{The corrected and normalised flux time series of NIK 1, in the region around OB1, OB2, and OB3. The SO is preceded by a NO, and 3 d after the SO we see a weaker, NO. There is a gap in the data covering the SO from BJD 2455182.5--2455185.4 due to the quarterly roll of the spacecraft and data downlink. \label{fig:soutburst}}
\end{figure*}

\begin{figure*}
\subfigure{\includegraphics[width=0.5\textwidth]{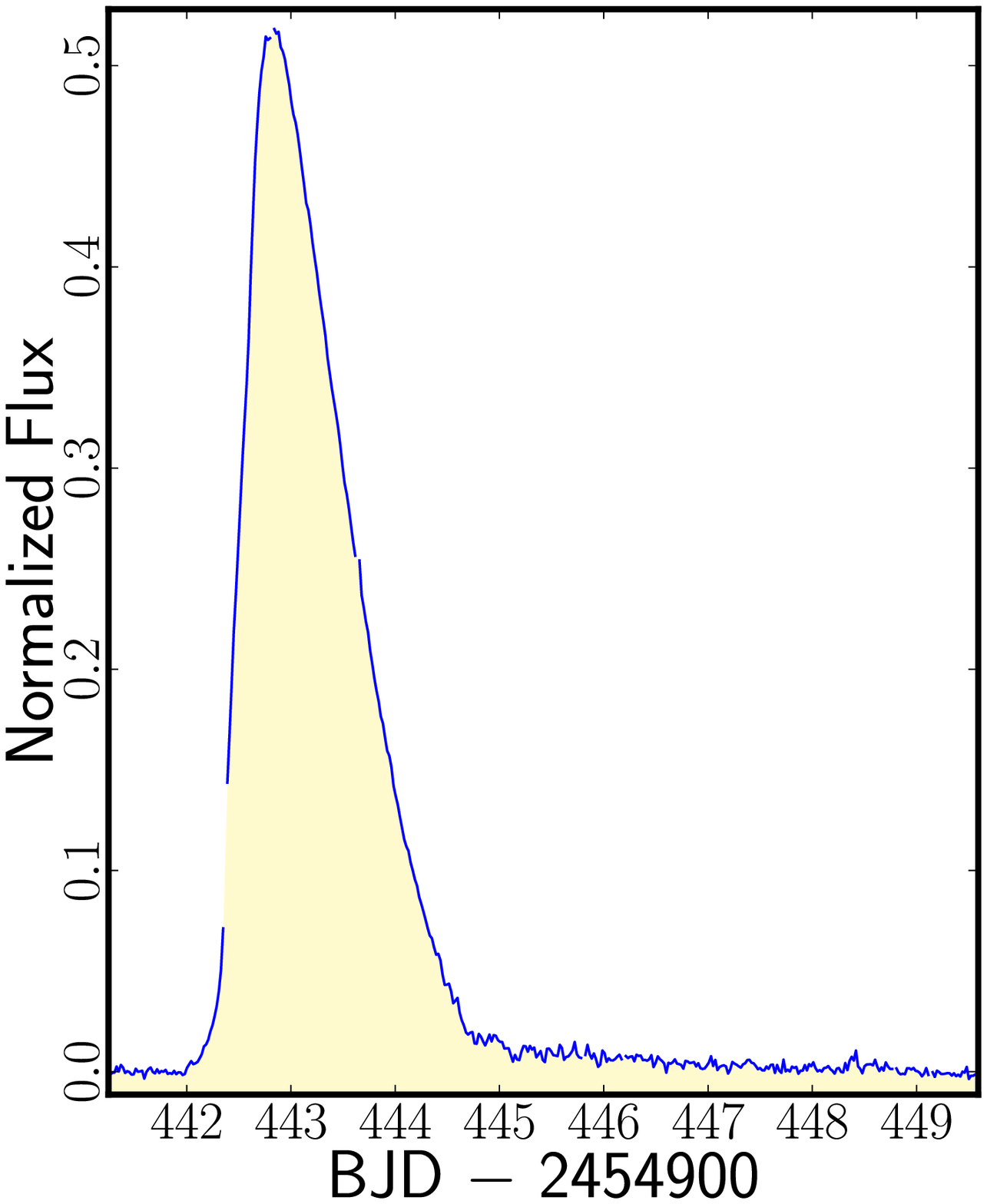}}\subfigure{\includegraphics[width=0.5\textwidth]{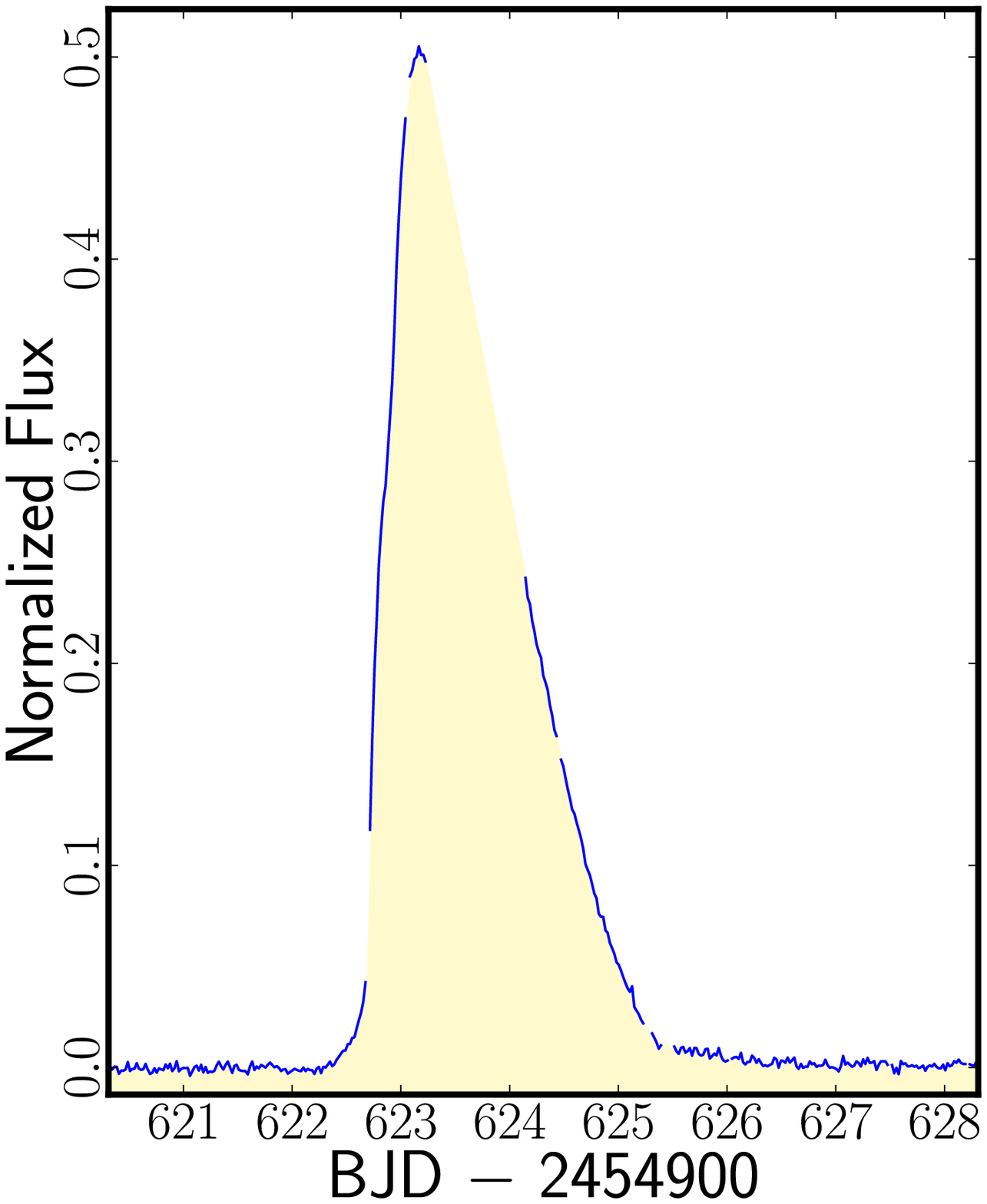}}
\caption{ The corrected and normalised flux time series of NIK 1, zoomed in on the region around OB4 in the left plot and OB5 in the right plot. The large data gap in the OB5 plot is owing to the monthly Earth-point occurring. No data is recorded during this time.\label{fig:noutburst}}
\end{figure*}

The four NOs are shown in detail in Fig.~\ref{fig:soutburst} and \ref{fig:noutburst}. OB4 and OB5 are very similar looking with comparable brightnesses. However, we caution that the uncertainty in measuring the outburst peak flux is high because of the method used to normalise these data. The shapes of the NOs are typical of those seen in other DN \citep[e.g. V344 Lyr,][]{cannizzo10} consisting of a fast rise to peak brightness of between 0.81--1.25 days with a somewhat slower fall to quiescence taking 1.86--2.57 days.

\subsection{The superoutburst}
The SO we detected, OB2, is shown in Fig.~\ref{fig:soutburst}. It reached peak flux on BJD 2455176.86 and returned to quiescence 10.42 days later. We note that the SO was immediately preceded by a NO, OB1, which did not return to quiescent brightness before the onset of the SO. OB3 then occurred only 4 days after the SO returned to the quiescent flux level.

\subsection{Superhumps}
Superhumps always occur during SOs \citep[e.g.][]{vogt80,warner95,kato11}. In NIK 1, the superhumps start during the decline of the NO, OB1, at the stage when the SO is likely beginning and continue until the OB3 begins.

We performed a Fourier transform of the SO searching for periods ranging from the Nyquist period (58.84 min) to 12 h. The strongest frequency ($F_{A}$) found is at $13.03\pm0.03$ cycles per day ($1.842\pm0.004$ h) where the uncertainty was calculated using a Monte Carlo technique \citep{lenz05}. In addition, we detect a signal at a frequency of $22.85\pm0.08$ cycles per day (=1.05 h, $F_{B}$). The first overtone of the strongest frequency should occur at 26.06 cycles per day ($F_{2}$). However, because the Nyquist frequency ($F_{N}$) of the data is 24.47 cycles per day, we cannot uniquely detect $F_{2}$. The value of $F_{2}-F_{N}$ is $1.59\pm0.06$ and $F_{N}-F_{B}$ is $1.62\pm0.08$, therefore, we concluded that the frequency at 22.85 cycles per day is very likely to be a reflection of the first overtone about the Nyquist frequency. 

\begin{figure*}
\includegraphics[width=\textwidth]{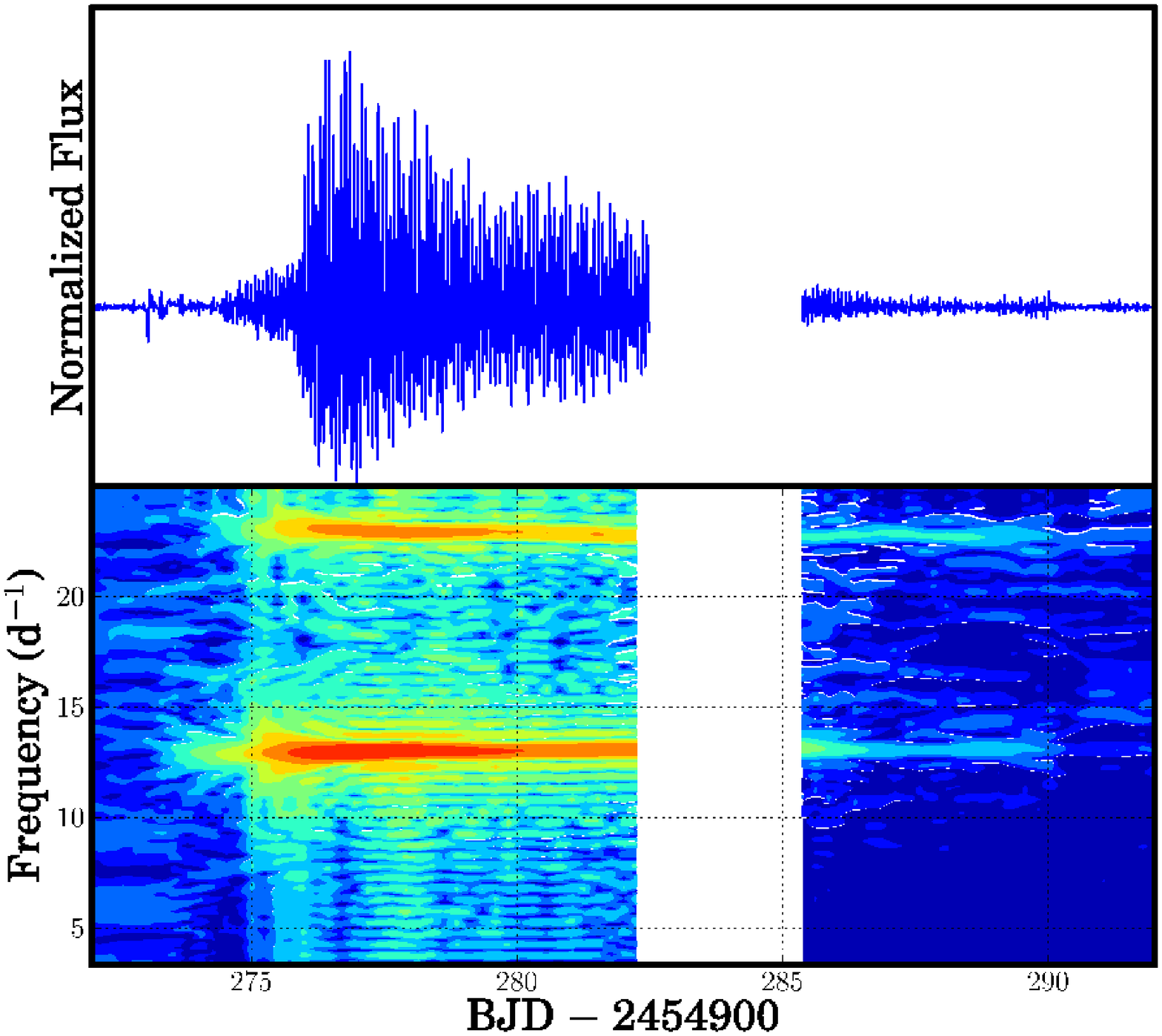}
\caption{The upper panel shows the SO after being propagated through a high-pass filter. The lower panel shows a Short Time Fourier Transform of the filtered data shown in the upper panel. Two periodic signals are detected: one at 1.842 h, the other at 1.05 h. These signals persist into the Q4 data and can be seen after the NIK 1 has faded before OB3 occurs. \label{fig:ftplot}}
\end{figure*}

In order to better characterise the superhumps, we propagated the time series through a boxcar high-pass filter with a cut-off frequency of 18.0 d$^{-1}$. This is shown in the top panel of Fig.~\ref{fig:ftplot}. We calculated a discreet Short Time Fourier Transform \citep{harris78} over the duration of the SO. We used the filtered data and calculated a Fourier transform every 0.1 d with a siding window size of 2.0 d. The resulting power spectrum is shown in the lower panel of Fig~\ref{fig:ftplot}. Both $F_{A}$ and $F_{B}$ are visible for the entire duration of the SO and persist through the interval between OB2 and OB3. The superhump falls below a $3\sigma$ detection level at BJD 2 455 190, close to the time OB3 begins. We do not detect the two signals over the rest of the time series with any significance.

\begin{figure*}
\includegraphics[width=\textwidth]{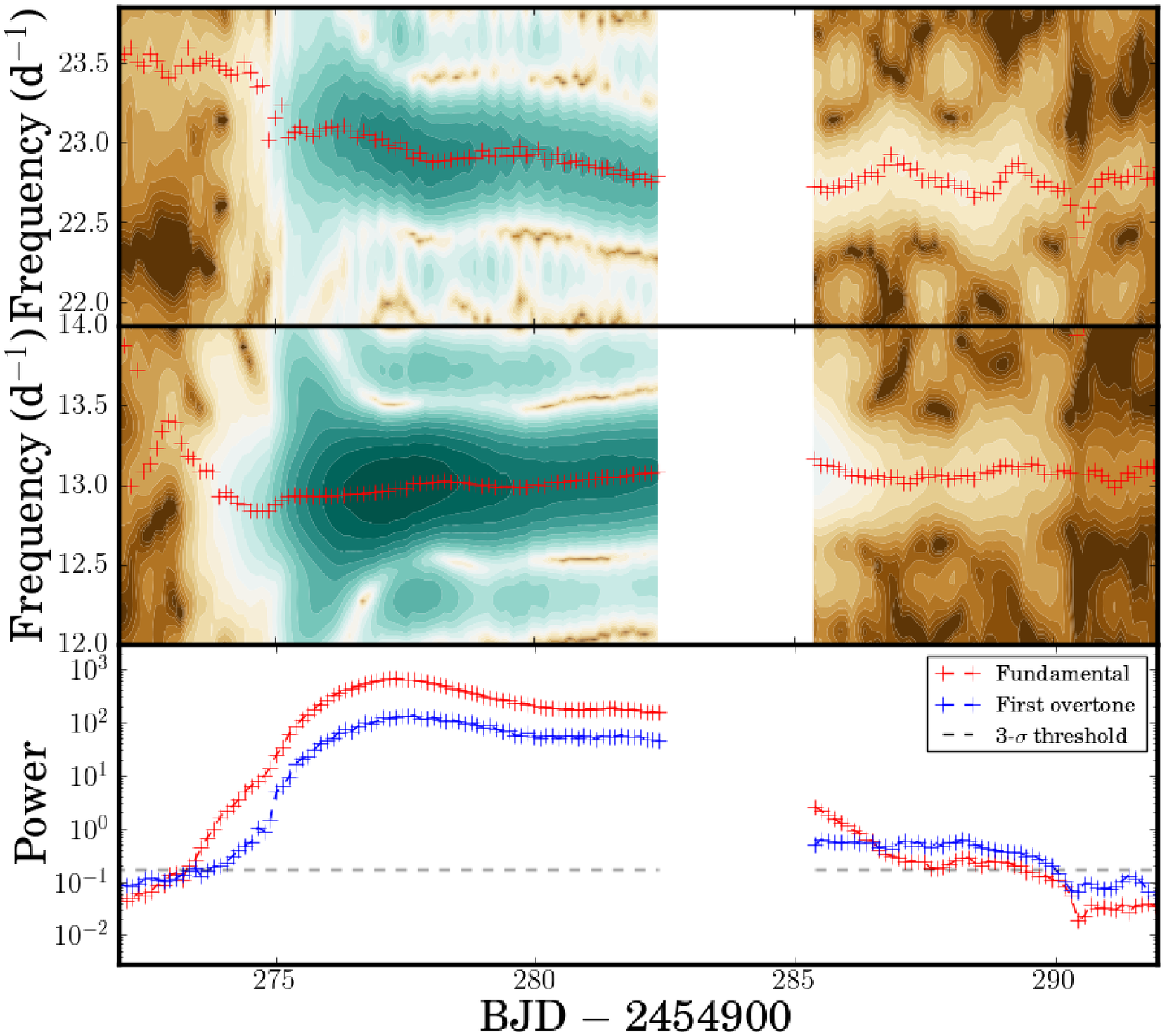}
\caption{The discreet Short Time Fourier Transform showing the two peaks identified in Fig.~\ref{fig:ftplot} in greater detail in the upper two panels. The red crosses show the position of the peak in the power spectrum in the region shown. The two panels' colors are scaled the same. The lower panel plots the peak power for both the fundamental frequency (red) and first overtone (blue). The black line is the 3-$\sigma$ detection threshold.
 \label{fig:closeft}}
\end{figure*}

In the upper two panels of  Fig.~\ref{fig:closeft} we zoom in around the two strongest frequencies seen in Fig.~\ref{fig:ftplot}. The peak frequency of both the fundamental and reflection of the first overtone frequency vary over time. The crosses show the position of the maximum power in the sliding window power spectrum. In Q3, the period of the main peak and the reflection of the frequency of the first overtone appear to move in opposite directions. This is expected if the higher frequency peak is indeed a reflection about the Nyquist frequency. In the lower panel of Fig.~\ref{fig:closeft} we show how the amplitude of both the fundamental  (red crosses) and the reflection of the first overtone (blue crosses) change over time. The second harmonic frequency decays slower than the fundamental. The fundamental frequency falls below the 3-$\sigma$ detection level 0.6 d before the first overtone frequency.

\begin{figure*}
\includegraphics[width=\textwidth]{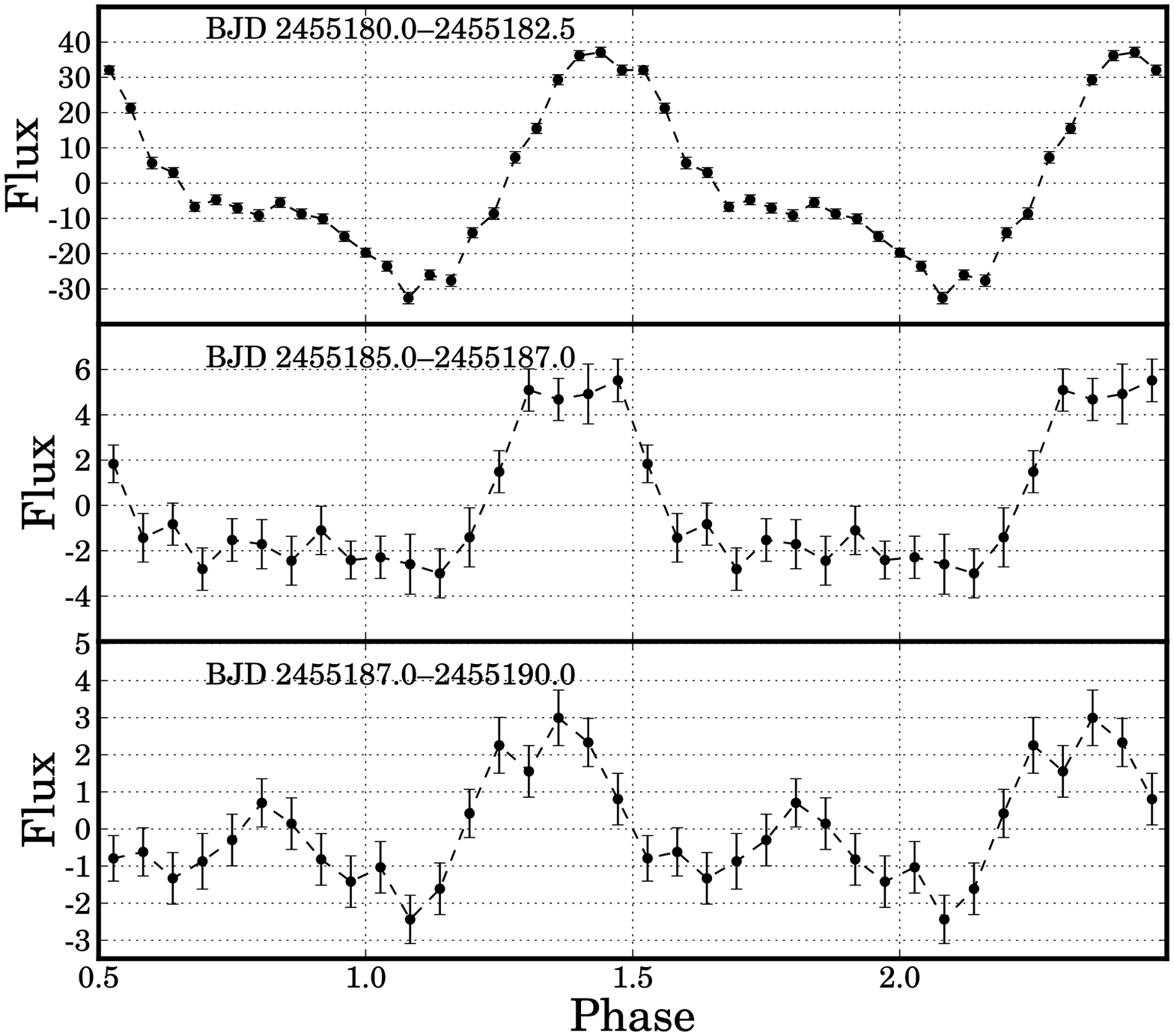}
\caption{Three sections of the light curve of NIK 1 folded on a period of 1.842 h and binned. The top panel covers the end of the last 2.5 days of the Q3 data, the middle panel the tail of the SO, which is the first two days of Q4 and the lower panel covers the quiescent period after the SO and rise of OB3. The flux levels are arbitrary but are internally consistent between the three plots.
 \label{fig:foldedlc}}
\end{figure*}

We see changes in the superhump waveform when comparing the modulation at mid-outburst (top panel in Fig.~\ref{fig:foldedlc}), at the tail end of the SO (middle panel), and when the system in a fainter state before the rise of OB3 (lower panel) . The third time series is double-peaked which explains the reflection of the second harmonic decaying slower than the fundamental frequency in Fig.~\ref{fig:closeft} --  the waveform now has twice the frequency of the initial waveform.

\section{Discussions}
\label{sec:disc} 


\subsection{The nature of the outbursts}

NIK 1 is now the fourth CV which has been seen in outburst by \K \citep{still10,kato11,garnavich11}. All four have undergone a SO. In two of the three other cases the SO has been immediately preceded by a NO \citep{still10,kato11}. In the third case, the rise of the SO is missed during a telescope safe-mode \citep{garnavich11}. While statistics are still low this raises a question: are all SOs triggered by a NO?

\citet{bateson77} suggested that in the DN VW Hyi, some SOs are triggered by a NO, and \citet{marino79} posited that this is the case for all SOs of VW Hyi. In a few other ground-based observations of DN, the SO has been observed to be preceded by a NO  -- e.g. T Leo \citep{kato97}, TV Crv \citep{uemura05} and GO Com \citep{imada04}. \citet{cannizzo10} postulates that a NO occurs due to thermal instability in the disc, and, when there is enough mass built-up from previous outbursts at large radial distances, a SO can be triggered. However, it takes time to propagate the heating front to large radii where this over-density of material in the outer disc exists. We first see the NO and then when outer material transitions to the hot state, the SO occurs.


The limit cycle disc instability model predicts the disc is much smaller directly after a SO and easier to excite into an outburst state. The intensity of outbursts should be lower and outbursts should occur more frequently for a smaller disc \citep{lasota01}. Our observation are consistent with this pattern. We observe no outbursts before OB1 -- the system was quiescent for 208 days -- and in the remaining observations the longest gap between outbursts is 181 days, the gap between OB4 and OB5, the final two outbursts. The NOs are more frequent after the SO and the spacing increases over time. We also see a much smaller outburst directly after the SO, which is consistent with theory.

We observed superhumps lasting for $\sim17$ d. They begin during the fading of the NO which preceded the SO and last until the beginning of OB3. We see superhumps for 3--4 days after the SO has faded to quiescence. There have only been four other reports of superhump persisting after the end of a SO, V1159 Ori \citep{patterson95}, ER UMa \citep{gao99}, EG Cnc \citep{patterson98}, and V344 Lyr \citep{still10}. It seems likely material will persist at large radii well after the majority of material has been accreted by the white dwarf and the reason superhumps have been seen persisting so few times is because of a lack of sensitivity in ground-based observations.


 The disc instability model \citep{osaki89,lasota01} predicts two main sources driving the superhump modulation; viscous dissipation within a flexing disc driven by the compression of the disc opposite the secondary star are single peaked, and the hot-spot where the accretion stream impacts the disc sweeping around the rim of a non-axisymmetric disc \citep{wood11}. During the \emph{early superhumps}, the hot-spot signal is much weaker than the signal from the disc and the phase curve is single peaked. As the disc fades, the hot-spot from the impact of the accretion stream onto the disc will become brighter relative to the disc because the disc has been drained of material and so is much less luminous. 

We see the sumperhump waveform change over the course of the SO. In the folded time series shown in  the upper panel of Fig.~\ref{fig:foldedlc} the waveform is asymmetric, the pulse rises around twice as fast as it falls. The shape is very similar to that seen in V344 Lyr \citep{still10} at a comparable time in the SO evolution and in the simulations shown in fig. 1 of \citet[][]{wood11} when modelling early superhumps.  In the middle panel, we see some evolution of the waveform and no longer see evidence for the asymmetry; the light curve is similar to that seen in fig. 3 of \citet[][]{wood11} where they have accretion stream impact in the model. Our assumption is that in the middle panel we are seeing no variable component from the accretion stream.


The waveform changes in the lower panel of Fig.~\ref{fig:foldedlc}. Similarly to the other panels there is a peak around phase 0.9 -- any difference can be attributed to the error on the period on which the data we folded. However, there is also a secondary peak which is offset in phase by $\sim180^{\circ}$ and of lower intensity than the main peak. This is wholly consistent with the fig. 2 of \citet[][]{wood11} where both the disc and the hot-spot contribute to the emission for the system.

Given that orbital periods are typically a few percent lower than the superhump period \citep{warner95}, we can place an upper limit on the binary orbital period of $1.842\pm0.004$ h, this places NIK 1 below the period gap. This meets expectations as almost all SU UMa stars are below the period gap \citep{kato11}.

\subsection{DN in the \K field of view}
What is the possibility of finding more DNe as background sources in the \K data? Using the empirical relation of \citet{patterson11}, the lowest peak absolute magnitude of a dwarf novae outburst in a system with a period longer than 90 min is $M_{V} = 5.3$. This implies \K can observe DNe in outburst out to a distance of 8.7 kpc if we assume we can observe all sources with a peak brightness $K_{P}<20$, which is approximately the source confusion limit of \emph{Kepler}. However, we cannot assume an isotropic distribution out to this distance. Instead, we restrict our search to only sources less than 300 pc above the Galactic plane, which \citet{patterson11} estimates to be the scale height for DNe.



The centre of the \K field of view is 13.5 deg above the Galactic plane and the field of view is 116 square degrees \citep{koch10}. Therefore the greatest distance an observable dwarf nova can be from Earth and still be within our scale height constraint is 1285 pc, and our search volume is $2.5\times 10^{7}$ pc$^{3}$ (this includes the entire field of view, not just the pixles downloaded). The period luminosity relation of \citet{patterson11} predicts all DNe outbursts in our search volume will be brighter than $V=15.8$ and easily be detected by \emph{Kepler}, even if they have high reddening.

\citet{patterson11} gives a lower limit of $1.8\times 10^{-6}$ DN pc$^{-3}$ but this value is calculated using only 42 systems with known distance and most likely under-predicts the true value significantly. Other estimates from observations include $7.9\times 10^{-6} \textrm{pc}^{-3}$ \citep{ak08}, $1.1\times 10^{-5} \textrm{pc}^{-3}$ \citep{pretorius07}, $2.1\times 10^{-5} \textrm{pc}^{-3}$ \citep{hertz90} and $10^{-4} \textrm{pc}^{-3}$ \citep{shara93}. Theoretically the number is thought to be around $1.8\times 10^{-4} \textrm{pc}^{-3}$ \citep{kolb93,pretorius07}. For each of these estimates we have calculated the inferred number of DNe potentially visible to \emph{Kepler}. These are given in Table~\ref{tab:spden} and assume an isotropic distribution within the search range of one Galactic disc scale height.

\citet{ak08} measures the scale height of the dwarf nova population to be only $150\pm18$ pc. If we use this value as the ceiling of our search the numbers predicted decrease by a factor of 8. These values are also shown in  Table~\ref{tab:spden}.

\begin{table*}
\caption{The number of dwarf novae predicted to be within one scale height of the Galaxy and in the \K field of view are given in this table. We use two estimates of the scale height; 300 pc and 150 pc. Likewise we use six different estimates of the space density which range from $1.8\times 10^{-6}$ to $10^{-4}$ pc$^{-3}$.}
\label{tab:spden}
\begin{tabular}{l|rrrr}
\hline
& \multicolumn{2}{|c|}{Sources in focal plane}&\multicolumn{2}{|c|}{Sources in downloaded data}\\
Space Density  & \multicolumn{2}{|c|}{Scale Height [pc]}&\multicolumn{2}{|c|}{Scale Height [pc]}\\
DN pc$^{-3}$& 300 (1)& 150 (2)&300 (1)&150 (2)\\
\hline
$1.8\times 10^{-6}$ (1)& 44.9&5.6&2.6&0.32\\
$7.9\times 10^{-6}$ (2)&196.8&24.6&11.3&1.4\\
$1.1\times 10^{-5}$ (3)&274.1&34.3&15.8&2.0\\
$2.1\times 10^{-5}$ (4)&523.4&65.4&30.1&3.8\\
$10^{-4} (5)$&2492&311.5&143.3&17.9\\
$1.8\times 10^{-4}$ (3,6)&4485&560.7&257.9&32.2\\
\hline
\multicolumn{5}{l}{(1) \citet{patterson11}, (2) \citet{ak08}, (3) \citet{pretorius07}}\\
\multicolumn{5}{l}{(4) \citet{hertz90}, (5) \citet{shara93}, (6) \citet{kolb93}}
\end{tabular}
\end{table*}

The predicted values range from 6--4485 which leads us to believe that the true space density of dwarf novae is wholly unconstrained.

If we are considering systems that may be found as background sources in apertures of target stars, the numbers decrease by a factor of $\sim17$ because \K only downloads the pixels from 5.75 per cent of the focal plane. With this restriction, the predicted number in \K ranges from 0.3--258. With only 9 known DNe in the data we are unable to rule out any of the models at the 3-$\sigma$ level. The predictions are likely to be an underestimate of the true number because this estimate would not include the source in this work as it would be further than 300 pc from the Galactic equator. A systematic search of the target pixel files may lead to a bounty of new DNe. Such work will help constrain the true value of the space density of DNe.

\section{Summary}
We have serendipitously observed outbursts of a DN which was captured within the aperture of a G-type star. We first extracted the flux of this source from the target pixel files and then removed systematic trends from the data using cotrending basis vectors. We observed four NO and one SO over 22 months. The SO was likely triggered by a NO. We observed superhumps during the SO and saw the waveform change from being single peaked to double peaked over the course of the outburst which we attribute to the appearance of the accretion stream impact onto the disc.

Given this DN was observed as a background source we used various predictions of DN space density to predict the number of DN in the \K field of view and in data already collected. The models vary wildly from predicting 4485 and 257.9 in the field of view and in collected data, respectively, to 2.6 and 0.32. A detailed study of the background pixels collected by \K could be used to constrain the space density of DNe.


\section*{Acknowledgments}
\K was selected as the tenth Discovery mission. Funding for \K is provided by NASA's Science Mission Directorate.
This paper made extensive use of the PyKE tools provided by the Kepler Guest Observer Office for the use of the community. 
All of the data presented in this paper were obtained from the Multimission Archive at the Space Telescope Science Institute (MAST). STScI is operated by the Association of Universities for Research in Astronomy, Inc., under NASA contract NAS5-26555. Support for MAST for non-HST data is provided by the NASA Office of Space Science via grant NNX09AF08G and by other grants and contracts. 
We used images from The Digitized Sky Surveys which were produced at the Space Telescope Science Institute under U.S. Government grant NAG W-2166. The images of these surveys are based on photographic data obtained using the Oschin Schmidt Telescope on Palomar Mountain and the UK Schmidt Telescope. The plates were processed into the present compressed digital form with the permission of these institutions.

%

\bibliographystyle{mn2e}
\bibliography{rats-new}





\end{document}